\documentclass{article}

\usepackage{arxiv}

\usepackage[utf8]{inputenc} 
\usepackage[T1]{fontenc}    
\usepackage{hyperref}       
\usepackage{url}            
\usepackage{booktabs}       
\usepackage{amsfonts}       
\usepackage{nicefrac}       
\usepackage{microtype}      
\usepackage{lipsum}		
\usepackage{graphicx}
\usepackage{natbib}
\usepackage{doi}
\usepackage{authblk} 
\usepackage{orcidlink} 
\usepackage{geometry}
\usepackage{multirow}%
\usepackage{amsmath,amssymb,amsfonts}%
\usepackage{amsthm}%
\usepackage{tikz}
\usepackage{mathrsfs}%
\usepackage[title]{appendix}%
\usepackage{xcolor}%
\usepackage{textcomp}%
\usepackage{booktabs}%
\usepackage{algorithm}%
\usepackage{algorithmicx}%
\usepackage{algpseudocode}%
\usepackage{listings}%
\usepackage{float}
\usepackage{array}
\usepackage{tabularx}
\usepackage{bm}
\usepackage{adjustbox}
\title{Explainable AI-based Decision Support for Nocturnal Hypoglycemia Prevention in Type 1 Diabetes}

\usepackage{makecell}    

\author{ \begin{tabular*}{\textwidth}{@{\extracolsep{\fill}}p{0.4\textwidth}p{0.3\textwidth}p{0.3\textwidth}} \parbox[c]{\linewidth}{\centering\textbf{Valentina Roquemen-Echeverri, PhD}\textsuperscript{1*}\\ \texttt{roquemev@ohsu.edu}} & \parbox[c]{\linewidth}{\centering\textbf{Peter G. Jacobs, PhD}\textsuperscript{1,2}\\ \texttt{jacobsp@ohsu.edu}} & \parbox[c]{\linewidth}{\centering\textbf{Leah M. Wilson, MD}\textsuperscript{3}\\ \texttt{wilsolea@ohsu.edu}} \\[2ex] \parbox[c]{\linewidth}{\centering\textbf{Joseph Pinsonault}\textsuperscript{1}\\ \texttt{pinsonau@ohsu.edu}} & \parbox[c]{\linewidth}{\centering\textbf{Deborah Branigan, BA}\textsuperscript{3}\\ \texttt{branigandebbie70@gmail.com}} & \parbox[c]{\linewidth}{\centering\textbf{Jae Eom, BSc}\textsuperscript{3}\\ \texttt{eom@ohsu.edu}} \\[2ex] \parbox[c]{\linewidth}{\centering\textbf{Daisy Chen, BSc}\textsuperscript{3}\\ \texttt{daisychen0@gmail.com}} & \parbox[c]{\linewidth}{\centering\textbf{Hantao Ling, PhD}\textsuperscript{3}\\ \texttt{hantao.ling@gmail.com}} & \parbox[c]{\linewidth}{\centering\textbf{Diana Aby-Daniel, PA}\textsuperscript{3}\\ \texttt{abid@ohsu.edu}} \\[2ex] \parbox[c]{\linewidth}{\centering\textbf{Kyle Chen}\textsuperscript{1}\\ \texttt{kylechen9948@gmail.com}} & \parbox[c]{\linewidth}{\centering\textbf{Clara Mosquera-Lopez, PhD}\textsuperscript{1}\\ \texttt{mosquera@ohsu.edu}} & {} \end{tabular*} }

\affil{\textsuperscript{*}Corresponding author}
\affil{\textsuperscript{1} Artificial Intelligence for Medical Systems (AIMS) Lab, Department of Biomedical Engineering, Oregon Health \& Science University (OHSU), 3303 S Bond Ave, Portland, 97239, Oregon, United States of America}

\affil{\textsuperscript{2}School of Chemical, Biological, and Environmental Engineering, Oregon State University, 1500 SW Jefferson Way, Corvallis, 97331, Oregon, United States of America}

\affil{\textsuperscript{3} Division of Endocrinology, Harold Schnitzer Diabetes Health Center, Oregon Health \& Science University (OHSU), 3303 S Bond Ave, Portland, 97239, Oregon, United States of America}

\renewcommand{\shorttitle}{Explainable AI-based Decision Support for NH Prevention in T1D}

\hypersetup{
pdftitle={A template for the arxiv style},
pdfsubject={q-bio.NC, q-bio.QM},
pdfauthor={David S.~Hippocampus, Elias D.~Striatum},
pdfkeywords={First keyword, Second keyword, More},
}

\begin{document}
\maketitle

\begin{abstract}

\textbf{Purpose:} Nocturnal hypoglycemia (NH) remains a challenge for individuals with type 1 diabetes (T1D), particularly those who are physically active or on multiple daily injections (MDI). We leveraged an explainable evidential neural network model that forecasts minimum overnight glucose to identify NH risk factors and generate recommendations for NH prevention. \textbf{Methods:} We analyzed the impact of glucose and physical activity (PA) factors on predicted NH probability using SHapley Additive exPlanations (SHAP). Data were from 20 adults with T1D on MDI (10 females; mean age 39 years; HbA1c 7\%) who participated in a cross-over study (NCT05967260). \textbf{Results:} A total of 502 nights were analyzed. Bedtime glucose was the strongest predictor of NH. Other relevant factors included recent exposure to high or low glucose, glucose variability before bedtime, and timing of PA. Some associations appeared physiologically counterintuitive, possibly reflecting behavioral adjustments. Based on the identified risk factors and their SHAP values, we determined key decision points and developed recommendations that may help prevent NH, such as consuming a bedtime snack or discussing potential adjustments to insulin therapy with a healthcare provider. \textbf{Conclusion:} Identifying predictors of NH offers insights for managing NH risk in clinical decision support.

\end{abstract}



\section{Introduction}\label{sec:introduction}

Nocturnal hypoglycemia (NH), defined as glucose $<70$ mg/dL during sleep, remains a major concern in type 1 diabetes (T1D), especially among individuals on multiple daily injections (MDI). Machine learning (ML) algorithms trained on data collected with wearables such as continuous glucose monitors (CGM), have the potential to improve management of T1D \cite{Jacobsen2023,Mallik2024}, and help prevent NH \cite{Bertachi2020,MosqueraLopezDodier2020,Jensen2020}. However, many ML models are “black boxes,” offering limited insight into how predictions are made to enable decision-making \cite{Yang2022,Kline2022}. This lack of transparency poses a challenge in clinical applications, where understanding the contribution of multi-modal health and behavioral data is essential for targeting desired glucose outcomes in T1D. Recently, several eXplainable Artificial Intelligence (XAI) techniques, such as SHapley Additive exPlanations (SHAP) \cite{Lundberg2017}, Local Interpretable Model-Agnostic Explanations \cite{Ribeiro2016}, and partial dependence analysis\cite{Friedman2001} have been developed. XAI have been increasingly applied in biomedical research \cite{Salih2024,Lotsch2022} as they provide insights into the relative contribution of inputs to predictions.

In this study, we leverage SHAP to interpret the NH predictions of an uncertainty-aware evidential neural network developed by our group \cite{MosqueraLopez2023Combining} and quantify the impact of glucose metrics and physical activity (PA) on the predicted probability of NH in T1D. Based on these associations, we identified key decision points and generated potential recommendations to reduce the likelihood of NH.

\section{Materials and methods}

This study is a secondary analysis of data collected from 20 free-living adults with T1D on MDI during the intervention arm of a single-center, open-label, randomized crossover study conducted between September 2023 and August 2024 (age 39±15 years; time since diagnosis 17±12 years; 10 females; 2 Asian, 2 Black/AA, 14 White/Non-Hispanic, 1 White/Hispanic, 1 more than one race; HbA1c 7.1±1.0\%). The primary aim of the study was to evaluate the feasibility and effect of a prediction-based bedtime snack in reducing NH. The study was conducted under US Food and Drug Administration approved investigational device exemption and the Oregon Health \& Science University Institutional Review Board approval. All participants provided written informed consent. The study is listed on ClinicalTrials.gov (NCT0596726) \cite{MosqueraLopez2025Evaluation}.

During the 4-week intervention period, CGM data were collected using the Dexcom G6 system (Dexcom Inc., San Diego, CA, USA). Participants self-reported PA through a survey completed every night before bed, which captured activity type, intensity, duration, and timing. These data were used as inputs to our previously developed uncertainty-aware evidential neural network model, the \textit{NHPredict} algorithm, that predicts the probability and timing of NH events (glucose values $<70$ mg/dL).  \textit{NHPredict} uses inputs from three domains: (1) glucose metrics derived from CGM data collected before bedtime; (2) PA type, duration, intensity, and timing; and (3) demographic factors, specifically age and biological sex. In prior evaluations, \textit{NHPredict} outperformed Elastic net, support vector regression, and XGBOOST models in terms of area under the receiver operating curve (AUROC) and other performance metrics, such as area under precision-recall curve (AUPRC) (See Table \ref{table:resultsAlgorith})  \cite{MosqueraLopez2023Combining}. The real-world clinical impact of \textit{NHPredict}, when combined with a digital health intervention, was subsequently evaluated through the clinical study described above, from where we obtained the data for this analysis.

\newcolumntype{C}[1]{>{\centering\arraybackslash}m{#1}}

\begin{table*}[b] 
\caption{Results of nocturnal hypoglycemia prediction methods in prior evaluations \cite{MosqueraLopez2023Combining}.} \small \resizebox{\textwidth}{!}
{%
\begin{tabular}{ C{4.0cm} C{3.0cm} C{1.8cm} C{2.5cm} C{1.8cm} C{4.2cm}} \toprule 
\multirow{2}{*}[-1em]{\makecell{\textbf{Nocturnal hypoglycemia}\\\textbf{time frame}}} & 
\multirow{2}{*}[-1em]{\makecell{\textbf{Performance}\\\textbf{metric}}} & 
\multicolumn{4}{c}{\textbf{Model}} \\ \cmidrule(lr){3-6} & & \makecell{\textbf{Elastic}\\\textbf{net}} & \makecell{\textbf{Support vector}\\\textbf{regression}} & \textbf{XGBoost} & \makecell{\textbf{Uncertainty-aware}\\ \textbf{evidential neural}\\ \textbf{network} \\(\textit{NHPredict})} \\ \midrule \multirow{2}{*}{\makecell{First half of the night\\0--4 h after bedtime}} & AUROC & 0.74 & 0.77 & 0.75 & \textbf{0.80} \\ & \makecell{AUPRC\\(chance level = 0.09)} & 0.32 & 0.38 & 0.32 & \textbf{0.43} \\ \midrule \multirow{2}{*}{\makecell{Latter half of the night\\4--8 h after bedtime}} & AUROC & 0.64 & 0.67 & 0.64 & \textbf{0.71} \\ & \makecell{AUPRC\\(chance level = 0.09)} & 0.17 & 0.20 & 0.18 & \textbf{0.22} \\ \bottomrule 
\end{tabular} 
} 

\label{table:resultsAlgorith} 
\end{table*}

To interpret the \textit{NHPredict}’s predictions, we used SHAP, a model-agnostic explainability framework grounded in cooperative game theory \cite{Lundberg2017}. SHAP assigns each model’s input a relevance value that represents its contribution to an individual prediction with respect to the average prediction across the training set. In our context, positive SHAP values indicated an increased predicted probability of NH, while negative values indicated a decreased probability.

\begin{figure}[!ht]
\centering
\includegraphics[width=\textwidth]{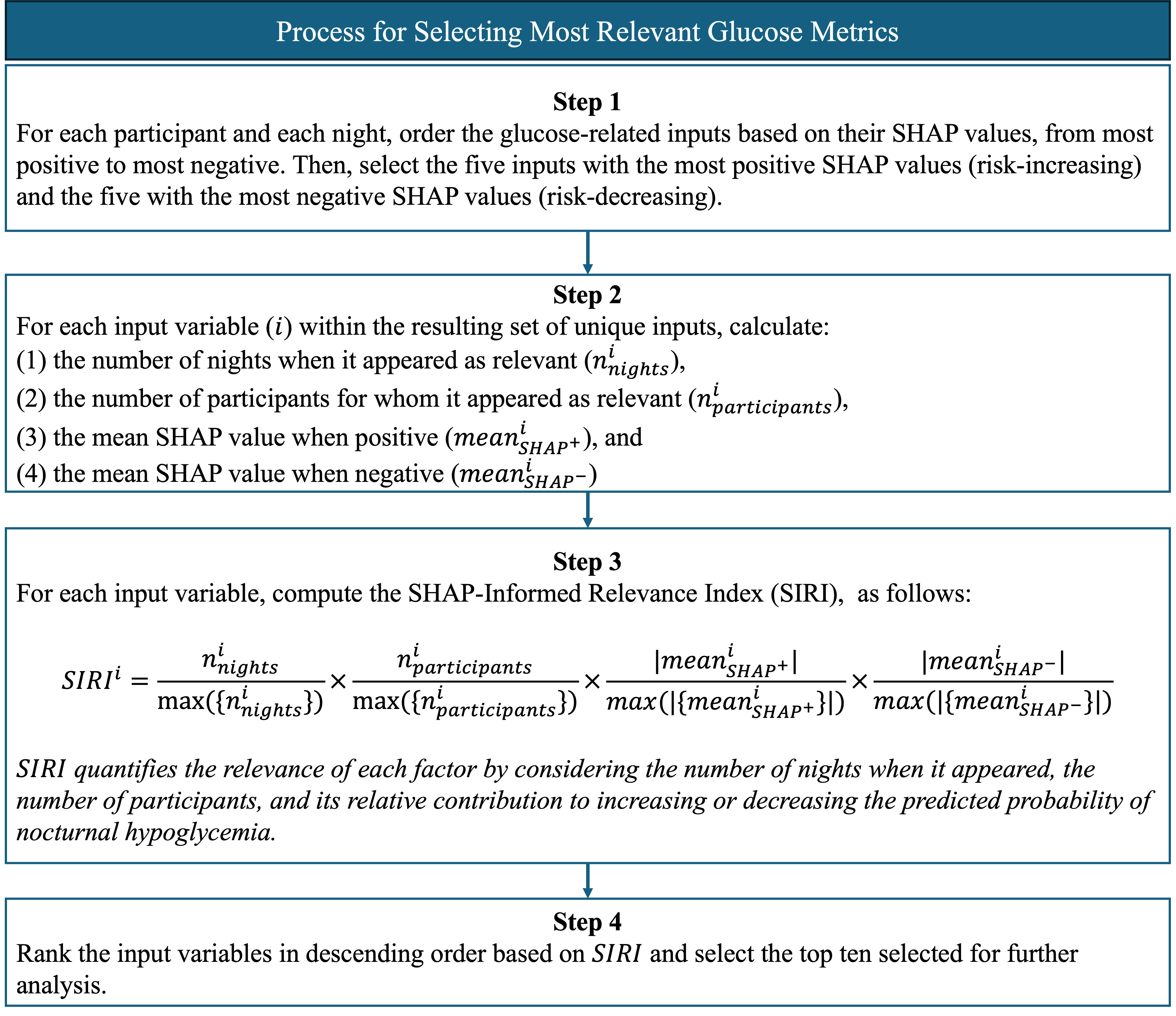}
\caption{Flow diagram of the process for selecting most informative glucose-related factors based on the SHAP-informed relevance index.}
\label{fig:NHfig1}
\end{figure}

For each night, we calculated the SHAP values for all input variables but focused on those related to the glucose and PA domains, as our goal was to analyze the impact of factors on which behavior modifications could be taken to prevent NH. We analyzed ten glucose-related factors that we identified as informative. The process for selecting these factors is shown in Figure \ref{fig:NHfig1}. Separately, when participants reported performing PA, we analyzed the SHAP values for the PA inputs to assess their influence on NH predictions. Since there are only four inputs related to PA, it was not necessary to apply the ranking process.

Once the SHAP values were computed for each glucose-related factor, we identified the factor value at which the SHAP value changed sign. These values were defined as thresholds, representing potential decision points at which interventions may be recommended to reduce NH risk. For each glucose-related factor, we used the identified threshold to propose a corresponding intervention aimed at lowering the likelihood of NH.

\section{Results and discussion}

\begin{figure}[!ht]
\centering
\includegraphics[width=0.85\textwidth]{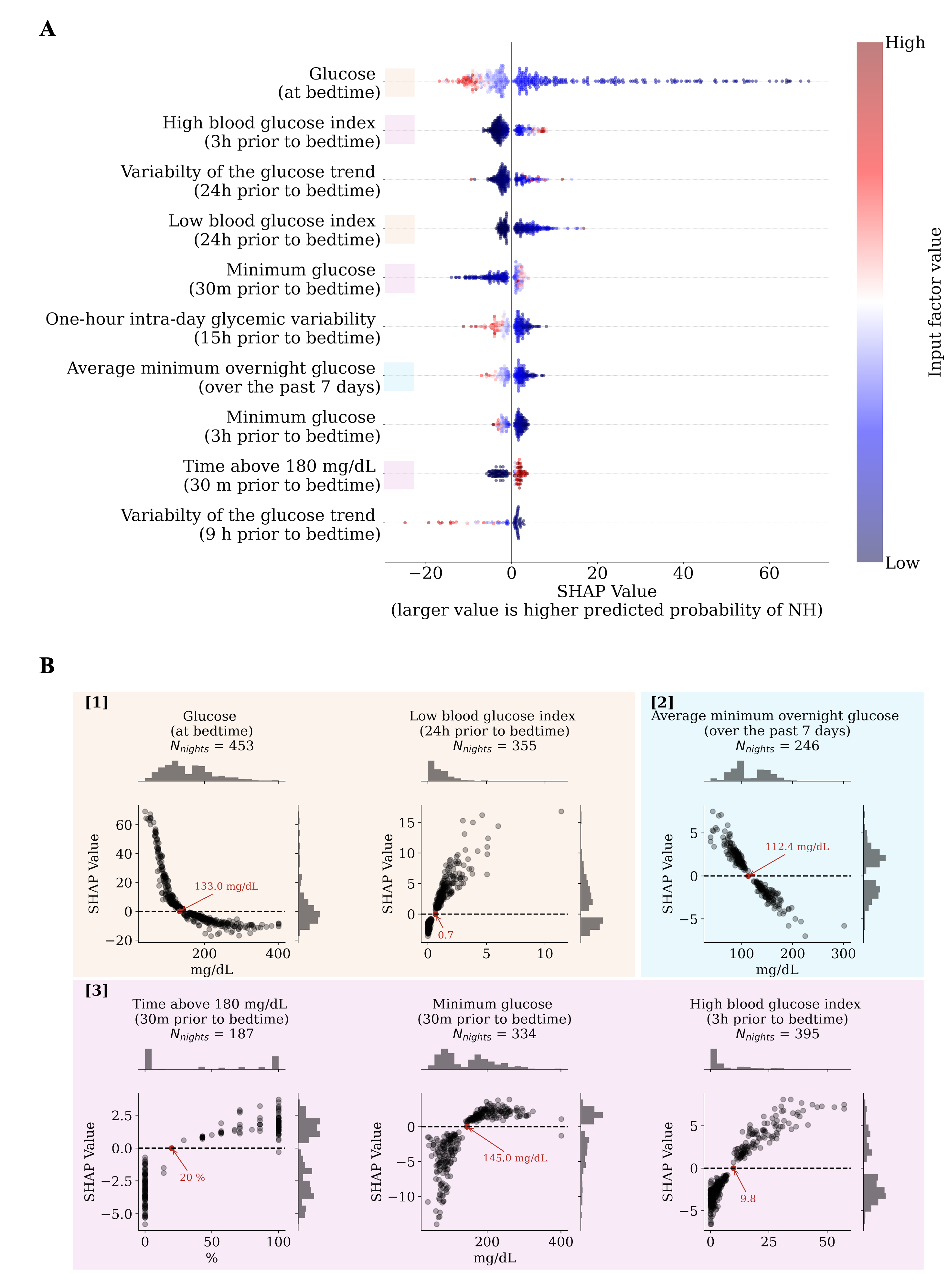}
\caption{\textbf{A.} The top ten glucose factors strongly associated with the predicted probability of nocturnal hypoglycemia. \textbf{B.} Distribution of SHAP values for glucose factors grouped by [1] glucose metrics previously identified as associated with NH, [2] glucose outcomes derived from the person’s historical overnight patterns associated with NH, and [3] glucose metrics that are associated with NH in a counterintuitive way. The number of nights ($N_{nights}$) in which each glucose factor ranked among the top five (risk-increasing) or bottom five (risk-decreasing) is shown. The red point indicates the zero-crossing SHAP value, which may serve as a threshold for decision-making aimed at reducing NH risk.}
\label{fig:NHfig2}
\end{figure}

A total of 502 nights were analyzed, and the top ten glucose-related factors associated with the predicted probability of NH are shown in Figure \ref{fig:NHfig2}A. The most significant factor was glucose at bedtime; lower values were associated with higher SHAP values, indicating an increased predicted probability of NH.

After qualitatively analyzing the top ten factors, we excluded the following from further analysis: variability of the glucose trend 24 hours prior to bedtime, one-hour intra-day glycemic variability 15 hours prior to bedtime, minimum glucose 3 hours prior to bedtime, and variability of the glucose trend 9 hours prior to bedtime because the time frames and interpretation of these factors are not easily actionable through decision support tools. The remaining six factors were grouped into three categories: [1] previously reported risk factors for nocturnal low glucose, [2] overnight CGM history, and [3] glucose metrics with counterintuitive associations with NH. Figure \ref{fig:NHfig2}B shows the SHAP value versus the feature value for each of the six selected factors grouped by the type of association between the factors and the predicted probability of NH.
Glucose at bedtime, along with the low blood glucose index during the 24 hours prior to bedtime, have been shown previously to be predictive of NH  \cite{MosqueraLopez2023ModelingRisk,Kovatchev1998,CoxGonder2007,Bisno2026}. This suggests that the \textit{NHPredict} is consistent with previously published findings (Figure \ref{fig:NHfig2}B[1])). Additionally, our analysis allowed us to quantify, for this population, the threshold at which a factor’s value shifts from reducing to increasing the predicted probability of NH, which can be used to provide decision support to help prevent NH. For example, when the bedtime glucose is higher than 133 mg/dL, this factor is no longer predictive of NH, and so a person could be encouraged to have a glucose above 133 mg/dL before bed. 

We also observed that the \textit{NHPredict} appears to identify how historical glucose patterns are predictive of NH (Figure \ref{fig:NHfig2}B[2]). For instance, when the average minimum overnight glucose over the prior 7 days is above 112 mg/dL, the factor is no longer predictive of NH. A decision support system could make use of this information to recommend actions (e.g. adjusting basal insulin delivery) to help a person avoid overnight low glucose. 

\begin{figure}[!ht]
\centering
\includegraphics[width=0.9\textwidth]{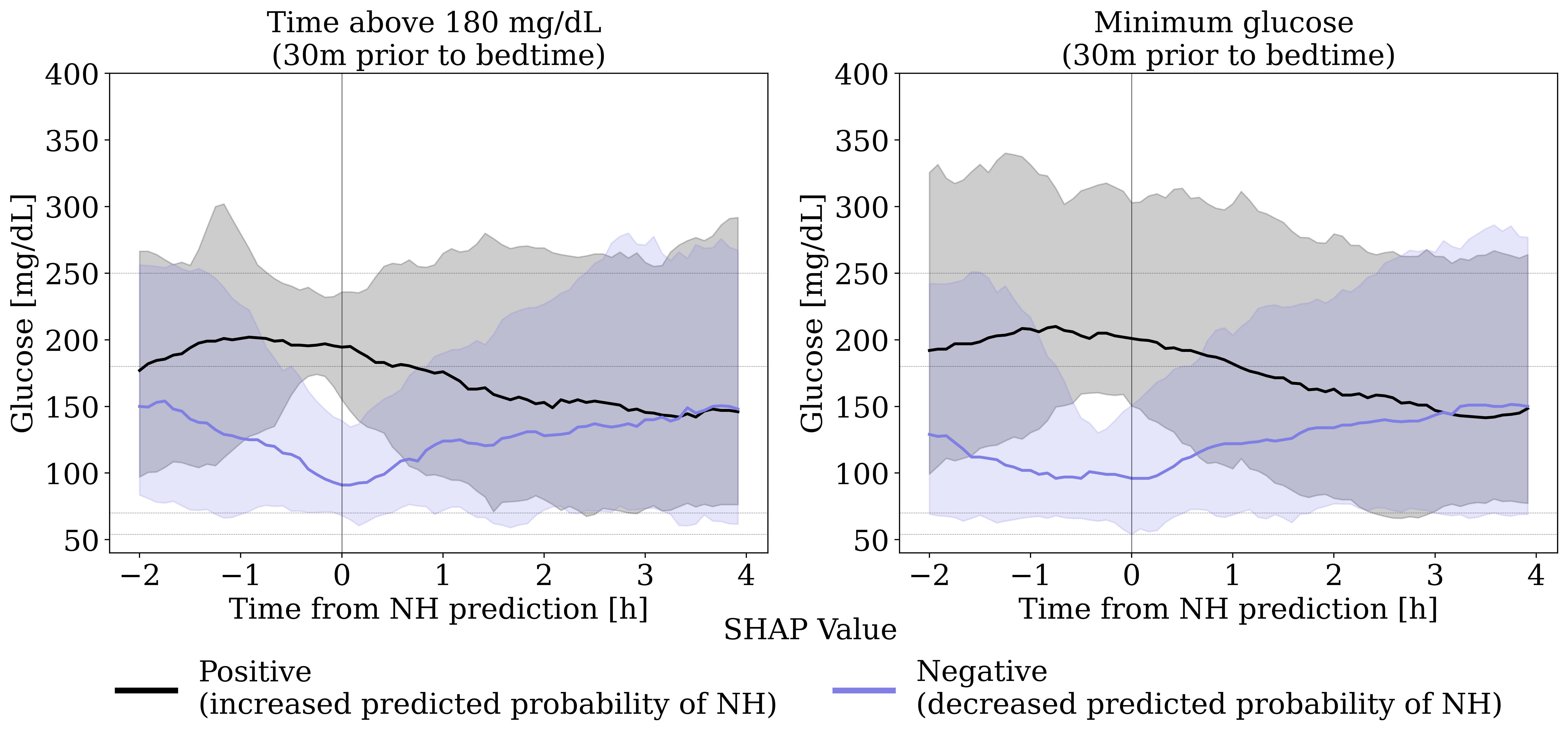}
\caption{CGM traces associated with physiologically counterintuitive SHAP values for glucose factors before bedtime. The solid lines correspond to the median glucose profile, and the shaded areas indicate the 5th to 95th percentile range.}
\label{fig:NHfig3}
\end{figure}

Some physiologically counterintuitive associations were also identified (Figure \ref{fig:NHfig2}B[3]). For example, a higher percentage of time spent above 180 mg/dL during the 30 minutes before bedtime was associated with an increased predicted probability of NH, while lower minimum glucose during the same time window was associated with a reduced probability. These associations may appear contradictory to the primary finding, that lower glucose at bedtime is generally associated with higher NH risk. However, they likely reflect behavioral patterns rather than the need of adjusting glucose thresholds. Specifically, such patterns may capture recent diabetes management actions, such as insulin corrections or bedtime snacks. We further investigated these counterintuitive factors by analyzing CGM traces on nights when these factors had either positive or negative SHAP values. Figure \ref{fig:NHfig3} shows these traces for time above 180 mg/dL before bedtime (left) and for minimum glucose before bedtime (right). In the former case, even when glucose levels were high before bed, SHAP values were positive, and CGM traces showed a decreasing trend in glucose after bedtime, likely due to an insulin injection, whose effects are usually delayed and therefore not captured by the glucose variables calculated at prediction time, potentially leading to NH. If our hypothesis is correct, this insight could be used to personalize the correction factor in this context. In the latter case, despite lower pre-bedtime glucose levels, an upward trend was observed in CGM data, likely reflecting a recent corrective action such as a meal or snack. These results show that SHAP is a useful tool for investigating the types of correlations that \textit{NHPredict} learns during training. Even when these relationships are counterintuitive, they were also observed in the data from our study. In the future, we plan to conduct a study that captures insulin and meal data to test whether our explanations of these correlations can be supported by objective data. 

\begin{figure}[!hb]
\centering
\includegraphics[width=0.8\textwidth]{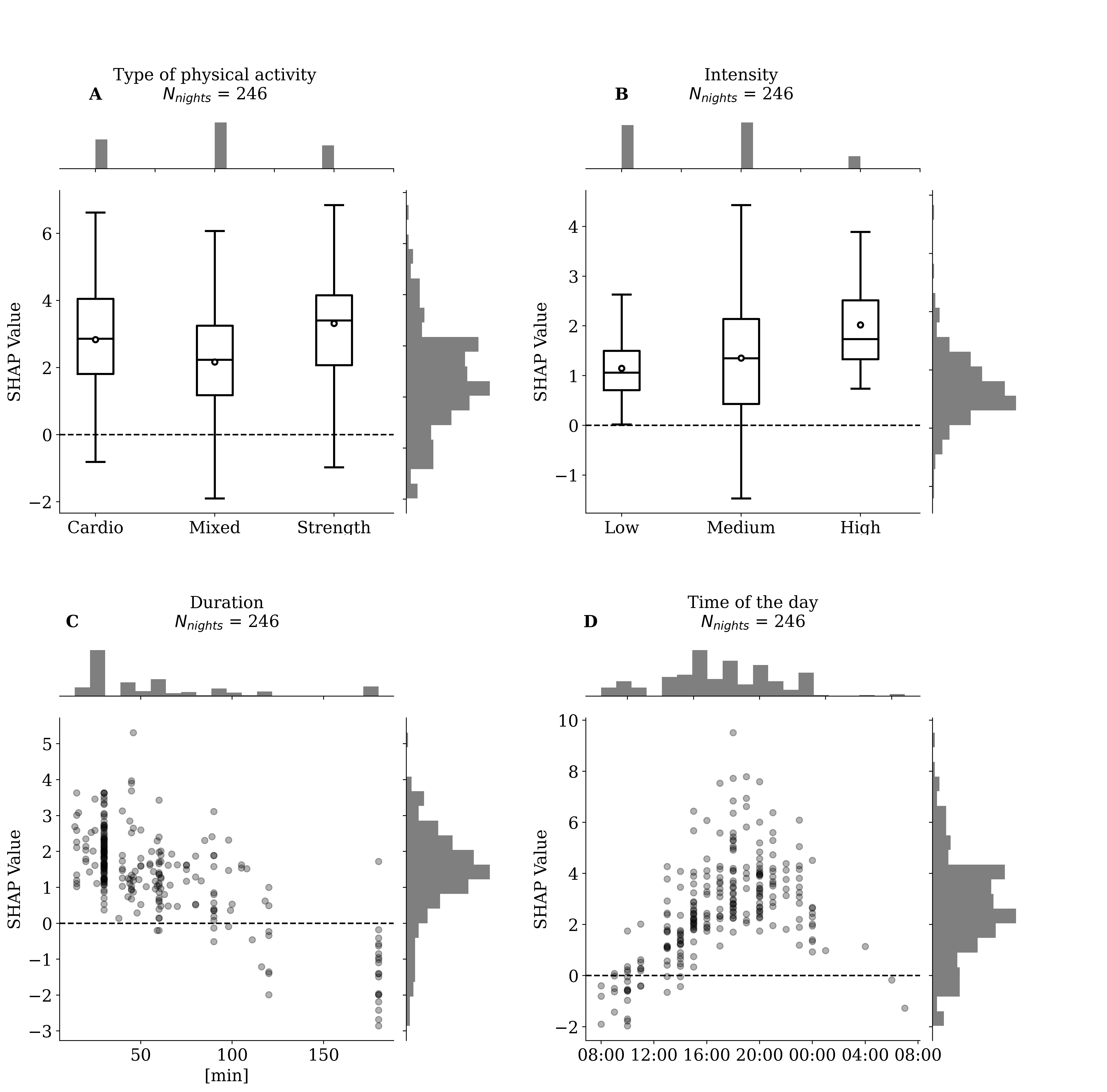}
\caption{Distribution of SHAP values for physical activity factors.}
\label{fig:NHfig4}
\end{figure}

Figure \ref{fig:NHfig4} presents the SHAP values for PA–related factors. A total of 246 nights included participant-reported PA. When comparing the median [IQR] of the distribution of SHAP values for days when participants did not report being physically active (0.05 [–0.10, –0.20]) versus days when they were physically active (1.09 [0.20,–2.50]), we found that PA was associated with an increased predicted probability of NH in line with previous reports \cite{MosqueraLopez2023ModelingRisk,Riddell2023,Wilson2015}. The SHAP coefficients do not appear to indicate that any of the exercise types were more predictive of NH (Figure \ref{fig:NHfig4}A) \cite{MosqueraLopez2023ModelingRisk}. We observed that exercising before midday (Figure \ref{fig:NHfig4}D) was associated with a reduced predicted probability of NH. This finding is also consistent with previously published work \cite{MosqueraLopez2023ModelingRisk,Riddell2023}.

Figure \ref{fig:NHfig4}C shows that NH risk decreases as exercise duration increases. However, we identified 18 reported physical activities with a duration of 3 hours, 13 of which came from the same participant. More data would need to be acquired to verify this relationship.

\begin{table*}[ht] 
\centering 
\caption{Recommendations for reducing nocturnal hypoglycemia (NH) risk based on the identified decision thresholds.\\[0.5ex] \textit{* This recommendation should not be applied when the individual is experiencing hyperglycemia or when other preventive measures to reduce the risk of NH have already been taken.}}
\small 
\renewcommand{\arraystretch}{1.4} \resizebox{\textwidth}{!}{%

\begin{tabular}{p{3.5cm} p{4.0cm} p{2.0cm} p{6.0cm}} 
\toprule 
\multicolumn{4}{c}{

\textbf{Glucose-related factors}} \\
\toprule 

\textbf{Glucose metric} & 
\textbf{Glucose metric interpretation}& 
\textbf{Triggering condition} & 
\textbf{Recommendation} \\ 

\toprule 

Glucose at bedtime & 
CGM glucose reading at bedtime & 
$<133$ mg/dL & 
\multirow{2}{6.0cm}[-1em]{Consume a bedtime snack to reduce the risk of overnight hypoglycemia.*} \\ 

Low Blood Glucose Index during the 24 hours prior to bedtime & 
Exposure to low glucose levels during the previous 24 hours & 
$>0.7$ & \\ 

\midrule 

Average minimum overnight glucose over the previous 7 nights & 
Average of the lowest overnight glucose values observed during the previous week & 
$<112$ mg/dL & 
Discuss basal insulin settings with a healthcare provider, as the basal insulin dose may be higher than required. \\ 

\midrule 

Time above 180 mg/dL during the 30 minutes prior to bedtime & 
Time spent in hyperglycemia immediately before bedtime & 
$>20\%$ & 
\multirow{3}{6.0cm}[-2em]{When correcting hyperglycemia before bedtime, consider a more conservative glucose target and discuss potential adjustments to correction factors or insulin dosing with a healthcare provider.} \\ 

Minimum glucose during the 30 minutes prior to bedtime & 
Lowest glucose value during the 30 minutes before bedtime & 
$>145$ mg/dL & \\ 

High Blood Glucose Index during the 3 hours prior to bedtime & 
Exposure to elevated glucose levels during the 3 hours before bedtime & 
$>9.8$ & \\

\toprule 
\multicolumn{4}{c}{\textbf{Physical activity-related factors}} \\
\toprule 
\multicolumn{3}{c}{\textbf{Triggering condition}} & 
\textbf{Recommendation} \\ 

\toprule 
\multicolumn{3}{c}{Late physical activity} & 
Consume a bedtime snack to reduce the risk of overnight hypoglycemia.* 

\\ \bottomrule 
\end{tabular}%
} 
\label{tab:recommendations} 
\end{table*}

Based on the identified decision thresholds, we translated the SHAP-derived associations into potential recommendations aimed at reducing the likelihood of NH. Table~\ref{tab:recommendations} summarizes the glucose- and physical activity-related factors, their user-facing interpretation, the triggering conditions, and the corresponding recommendations. For factors previously associated with NH, such as bedtime glucose and exposure to low glucose during the preceding 24 hours, the recommendations primarily focus on preventive strategies such as consuming a bedtime snack. For factors reflecting longer-term overnight glucose patterns, recommendations emphasize discussing potential adjustments to basal insulin therapy with a healthcare provider. Finally, for factors showing counterintuitive associations with NH, the recommendations focus on more cautious management of pre-bedtime hyperglycemia. These recommendations are intended to illustrate how explainable machine learning can help identify actionable decision points for clinical decision support; however, they should not be interpreted as standalone clinical recommendations, as they do not account for interactions among variables or the broader clinical context.

\section{Conclusion}

SHAP values provide insights into which inputs variables in a “black box” ML model are most important when making predictions, enabling informed clinical decision-making and providing tools to help educate people with T1D to recognize and reduce high-risk situations and behaviors. The \textit{NHPredict} aligns with scientific literature and learns logical patterns from the data to predict the probability of NH, although these patterns are not necessarily causal and, in some cases, might be physiologically counterintuitive. This suggests additional context might be required to better interpret some of the \textit{NHPredict}’s predictions.

We identified thresholds for glucose- and physical activity-related factors that were associated with an increased predicted probability of NH. These thresholds were used to generate potential recommendations, ranging from consuming a bedtime snack to discussing adjustments to insulin therapy parameters including basal dose and correction factors with a healthcare provider. Together, these findings demonstrate how explainable ML methods can be used to transform NH risk predictions into actionable decision-support recommendations for individuals living with T1D.

\subsection*{Acknowledgements}

We would like to thank our study participants for dedicating their time to support diabetes research and Dexcom Inc. for providing Dexcom G6 sensors and transmitters for this study.

\subsection*{Competing Interests}

P.G.J. reports advisory board participation and research support from Eli Lilly and Dexcom Inc. P.G.J. have a financial interest in Pacific Diabetes Technologies, a company that might have a commercial interest in the results of this research and technology. L.M.W. received research support from Eli Lilly and Dexcom Inc. and is now employed by Dexcom Inc. C.M-L is employed by Insulet Corporation. All other authors declare no conflicts of interest.

\subsection*{Funding}

This study was supported by the National Institute of Diabetes and Digestive and Kidney Diseases (grant NIH/NIDDK 5R21DK128582 to C.M-L.).


\bibliographystyle{unsrt}
\bibliography{references}  

\end{document}